\documentclass[aps, prl, amsmath, amssymb, superscriptaddress, a4paper, floatfix]{revtex4-2}

\usepackage{graphicx} 
\usepackage[utf8x]{inputenc}
\usepackage{physics}
\usepackage{siunitx}
\usepackage{textgreek}
\usepackage[version=4]{mhchem}
\usepackage{comment}

\begin{document}
\title{Positive pressure matters in acoustic droplet vaporization}

\author{Samuele Fiorini}
\author{Anunay Prasanna}
\author{Gazendra Shakya}\author{Marco Cattaneo}
\author{Outi Supponen}

\affiliation{Institute of Fluid Dynamics, ETH Zürich, Zürich, Switzerland}

\date{\today}

\begin{abstract}
    Acoustically vaporizable droplets are phase-change agents that can improve the effectiveness of ultrasound-based therapies. In this study, we demonstrate that the compression part of an acoustic wave can generate tension that initiates the vaporization. This counter-intuitive process is explained by the occurrence of Gouy phase shift due to the focusing of the acoustic wave inside the droplet. Our analysis unifies the existing theories for acoustic droplet vaporization under a single framework and is supported by experiments and simulations. We use our theory to identify governing parameters that allow to vaporize droplets using predominantly compression waves, which are safer in medical use.
\end{abstract}

\maketitle

\emph{Introduction.}
Micron- and sub-micron-sized superheated liquid perfluorocarbon droplets are a relatively new class of ultrasound-responsive agents that are typically metastable in blood circulation and can be converted into microbubbles upon acoustic excitation~\cite{kripfgans_acoustic_2000}, with the possibility to target specific areas of the human body.
This process, commonly known as acoustic droplet vaporization (ADV), shows promise in applications such as contrast-enhanced ultrasound imaging~\cite{sehgal_sonographic_1995}, gas embolotherapy~\cite{harmon_minimally_2019}, targeted drug delivery~\cite{chen_targeted_2013}, and other ablation techniques~\cite{vlaisavljevich_effects_2015, zhang_acoustic_2011}. Compared to microbubble contrast agents, droplet emulsions present greater stability against dissolution in the blood flow \cite{borden_acoustic_2020} and can be nanometric in size to leverage the Enhanced Permeability and Retention (EPR) effect to improve drug accumulation in tumor tissue \cite{maeda_epr_2013}.
Vaporization inception is typically characterized by an acoustic threshold, which depends non-trivially on parameters such as droplet radius, frequency, and temperature~\cite{shakya_ultrasound-responsive_2024}.

Significant progress has been made in elucidating the physics of ADV~\cite{shpak_ultrafast_2013,  reznik_efficiency_2013, shpak_acoustic_2014, li_initial_2014, qin_predicting_2021}, as well as in characterizing and reducing the acoustic vaporization threshold \cite{kripfgans_acoustic_2000, aliabouzar_acoustic_2018, aliabouzar_effects_2019}, which currently greatly exceeds the pressure required to excite microbubble contrast agents \cite{wu_investigation_2021} and can question the safety of droplet-based \emph{in vivo} treatments. 
Shpak \emph{et al.}~\cite{shpak_acoustic_2014} suggested that the focusing of superharmonics of the driving wave, created due to nonlinear propagation in the medium surrounding the droplets, can generate regions of localized tension amplification in the liquid core that facilitate nucleation. They experimentally showed that this effect is particularly pronounced for large micrometric droplets (radius $ > 4 $ \si{\micro\meter}), and further supported their results by means of theoretical modeling.
Lajoinie \emph{et al.}~\cite{lajoinie_high-frequency_2021} described another possible mechanism for droplet vaporization based on acoustic resonance. By solving a linearized equation for the droplet radial dynamics, they theoretically predicted the resonance frequency and the pressure amplification factor. 
The calculated maximum amplification factor for a resonant pressure field is practically constant with respect to droplet radius, making acoustic resonance a good candidate to exploit for vaporizing sub-micron droplets. 

While most of the prior studies on ADV focus on the peak negative pressure (PNP) as the main contributor to vaporization initiation, in this Letter we demonstrate that the compressive phase of the driving acoustic wave can play a significant role in triggering phase-change by means of a physical mechanism that has been previously overlooked. Specifically, it can dictate the location of the maximum tension within the droplet bulk due to the occurrence of \emph{Gouy Phase Shift}~\cite{gouy_sur_1890}, which leads to a sign reversal of the focusing incoming wave as it crosses the focal point. This new insight advances our understanding of the physics behind ADV and the presented analysis unifies the superharmonic focusing and resonance theories under the same theoretical framework.

\emph{Theory.}
We model the axisymmetric acoustic wave propagation within a spherical droplet of radius $R$,  density $\rho_2$, and speed of sound $c_2$, located in an infinite medium of density $\rho_1$ and speed of sound $c_1$ as described elsewhere \cite{shpak_acoustic_2014, anderson_sound_1950, feuillade_anderson_1999}. The transmitted pressure within the droplet is obtained by summing the solutions for every harmonic component, $n$, of the incoming nonlinear ultrasound wave with fundamental angular frequency $\omega$ and initial phase $\phi$, owing to the linearity of the wave equation. 
The pressure field in the droplet core can be described as a function of the radial coordinate $r$, the polar angle $\theta$, and the time $t$ as:
\begin{equation}
    \resizebox{0.5\textwidth}{!}{$p_{t}(r, \theta, t) = \Re \left[ \sum_{n = 0}^{\infty} \sum_{m = 0}^{\infty} a_ne^{in\omega t + \phi_n}\tilde{\alpha}_{m,n}j_m(nk_2r)P_m(\cos{\theta})\right]$},
    \label{eq:transmittedPressure}
\end{equation}
where
\begin{equation}
    \resizebox{0.5\textwidth}{!}{$\tilde{\alpha}_{m,n} = (-i)^m(2m + 1)\frac{j_m(nk_1R)h_m^{\prime(2)}(nk_1R) - h_m^{(2)}(nk_1R)j_m^{\prime}(nk_1R)}{j_m(nk_2R)h_m^{\prime(2)}(nk_1R) - \frac{\rho_1c_1}{\rho_2c_2}h_m^{(2)}(nk_1R)j_m^{\prime}(nk_2R)}$},
    \label{eq:amplficationFactors}
\end{equation}
is the amplification factor with $k_1 = \omega/c_1$ the fundamental wavenumber in the surrounding medium, $k_2 = \omega/c_2$ the fundamental wavenumber inside the droplet, $j_m$ and $y_m$ the spherical Bessel functions of the first and second kind of order $m$, $h_m = j_m - iy_m$ the spherical Hankel function of the second kind of order $m$ and the prime represents differentiation with respect to $k_q r$, where $q = 1,2$. The complete derivation is provided in the Supplementary Material \cite{noauthor_notitle_nodate}.

\begin{figure}[tb]
    \includegraphics[width = 0.5\columnwidth]{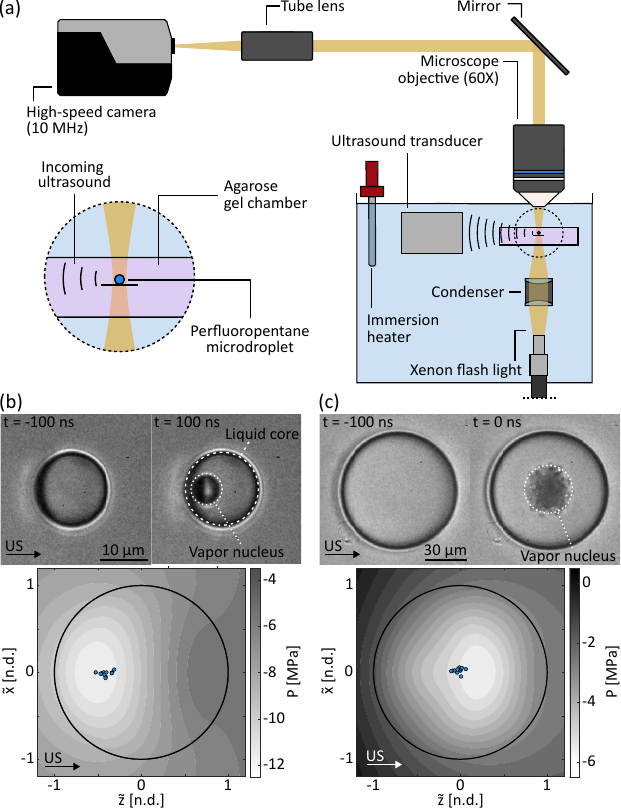}
     \caption{\label{fig:setupPressureMap} (a)~Schematic of the experimental setup. (b)~\emph{Top}: Snapshots of the vaporization dynamics of a single 9-\si{\micro\meter}-radius droplet sonicated with a 5-\si{\mega\hertz} HIFU. \emph{Below}: Experimentally observed nucleation positions (blue dots) overlapped with the numerically simulated instantaneous pressure field on the droplet center plane when the global minimum pressure is reached. (c)~Similar to (b), with a 38-\si{\micro\meter}-radius droplet and a 3.5-\si{\mega\hertz} HIFU.
     In both cases, nucleation inception happens at the time $t = 0 \; \si{\nano\second}$.}
\end{figure}

\emph{Numerical simulations.}~Two-dimensional axisymmetric numerical simulations are performed using the k-Wave toolbox in MATLAB \cite{treeby_k-wave_2010}. 
The experimentally measured pressure wave at the transducer focal point is used to calculate the temporal profile of a velocity source in the $z$-direction and to create a traveling plane wave in the simulation domain.
The droplet is modeled as a circular discontinuity in density and sound speed, placed in an infinite homogeneous medium (water). The simulated pressure fields on the droplet centerline show excellent agreement with the theory upon reaching steady-state, thereby validating the theoretical model~(see Supplementary Material for  details \cite{noauthor_notitle_nodate}).

\emph{Experimental setup.}~Monodisperse perfluoropentane (\ce{C5F12}, PFP, ABCR Swiss AG) droplets with an average radius of 9 \si{\micro\meter} and 38 \si{\micro\meter} are prepared using polydimethylsiloxane (Elastosil RT601 A/B, Wacker Chemie AG) microfluidic chips fabricated in-house following the protocols described elsewhere \cite{bardin_high-speed_2011, liu_generation_2018}.
A 5\% v/v solution of fluorosurfactant (Capstone\texttrademark{} FS-30, Apollo Scientific) is employed as the continuous phase to ensure the long-term stability of the droplets. 
Polydisperse droplet populations are prepared by emulsification of 1\% v/v fluorosurfactant solution and perfluoropentane using an amalgamator (MHC Technology, FR).

The experimental setup employed to characterize ADV is illustrated in Fig.~\ref{fig:setupPressureMap}(a).
Vaporization dynamics are visualized using a custom-built vertical microscope~\cite{cattaneo_shell_2023} with a 3-\si{\milli\meter} focal length objective (Olympus LumPlanFL N 60x water dipping) and the possibility to mount either a 600-\si{\milli\meter} focal length (TL600-A, Thorlabs) or a 200-\si{\milli\meter} focal length (TTL200-A, Thorlabs) tube lens, providing an overall magnification of 200$\times$ and 66.66$\times$, respectively. 
The microscope is coupled to a high-speed camera (Shimadzu HPV-X2) set to record at 10 million frames per second using backlight illumination provided by two xenon flash lamps used sequentially (MVS-7010, PerkinElmer).
The droplets are placed within a 1\% w/v agarose test chamber, submerged in a water tank maintained at an ambient temperature of 37--40~\si{\celsius} by an immersion heater.
The droplets are subjected to acoustic excitation from a 5-\si{\mega\hertz} (Precision Acoustics, UK) and 3.5-\si{\mega\hertz} (Olympus, JPN) single-element high-intensity focused ultrasound (HIFU) transducer, driven by an arbitrary wave generator (LW420B, Teledyne LeCroy) and amplified by a 53-\si{\decibel} linear power amplifier (1020L, E\&I). A linear array probe (LA332, Esaote, IT) controlled by an ultrasound advanced open platform (ULA-OP, MSD Lab, Università degli Studi di Firenze, IT \cite{tortoli_ula-op_2009}) is additionally tested to investigate the influence of wave distortion on the acoustic threshold.
The pressure field at the optical and acoustic focus is experimentally measured using a 75-\si{\micro\meter} needle hydrophone (NH0075, Precision Acoustics, UK). 

\emph{Results and discussion.}~Experimental frames depicting the initial stage of ADV are presented in Fig.~\ref{fig:setupPressureMap}(b) for a 9-\si{\micro\meter} and Fig.~\ref{fig:setupPressureMap}(c) for a 38-\si{\micro\meter}-radius droplet upon excitation with a 5-\si{\mega\hertz} and 3.5-\si{\mega\hertz} HIFU, respectively (the full recordings are available in the Supplementary Material~\cite{noauthor_notitle_nodate}).
For both recordings, $ t = 0 $ \si{\nano\second} represents the time when the first nucleated vapor bubble becomes visible.
The numerically computed instantaneous pressure field corresponding to the global pressure minimum is shown below the images and compared with the experimentally observed nucleation locations. 
As expected, the nucleation spots are positioned close to the region of global minimum pressure, which in these specific conditions (acoustic wave, droplet radius, temperature) is highly localized either within the proximal side of the droplet or close to its center.

\begin{figure}[tb]
   \includegraphics[width = 0.5\columnwidth]{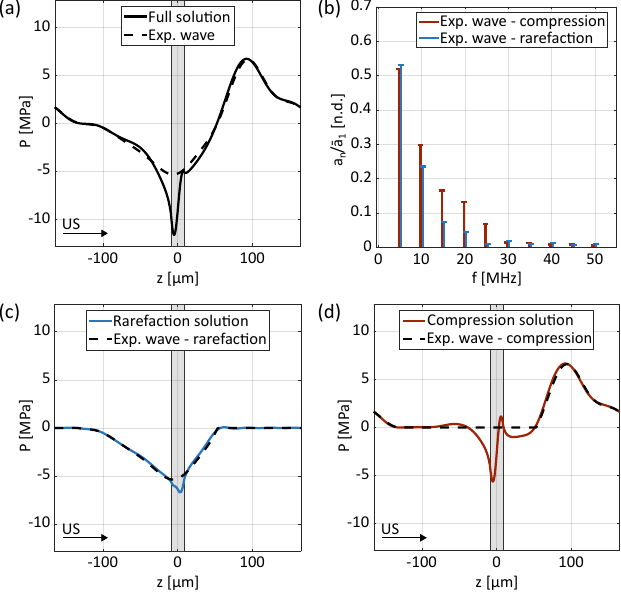}
   \caption{\label{fig:CompressionVsRarefaction} (a)~Theoretically calculated pressure field along the symmetry axis of a 9-\si{\micro\meter} radius droplet, compared with a single cycle of the free-field, experimentally measured 5-\si{\mega\hertz} HIFU pressure signal. (b)~Harmonic content of the compression and rarefaction phase of the experimentally measured pressure wave, normalized by the coefficient of the fundamental harmonic of the complete wave. 
   (c)~Transmission of the rarefaction and (d)~compression part of the experimentally recorded wave computed on the droplet centerline compared with the respective undisturbed reference. The shaded area represents the droplet position.
   All snapshots depict the time instant at which the minimum pressure of the full solution is reached.}
\end{figure}

The free-field experimental pressure profile generated by the 5-\si{\mega\hertz} HIFU, used as an input for the theoretical model, is depicted in Fig.~\ref{fig:CompressionVsRarefaction}(a) together with the theoretically evaluated pressure field along the centerline of the 9-\si{\micro\meter}-radius droplet.
The experimental pressure wave presents a marked asymmetry between the positive and negative peaks due to the wave's distortion upon nonlinear propagation in water and the resulting presence of superharmonics.
To investigate the effect of the distortion on tension generation in the droplet core, the wave is split into its compression and rarefaction parts, and the harmonic content of the resulting monopolar waves, shown in Fig.~\ref{fig:CompressionVsRarefaction}(b), is then used to compute two separate pressure fields.
Similar to examining distinct harmonic components of a wave, the linearity of the wave equation implies that the complete pressure field within the droplet core can be recovered by summing these two partial contributions.

Figs.~\ref{fig:CompressionVsRarefaction}(c)-(d) display two instantaneous pressure profiles corresponding to the monopolar wave solutions at the instant of time in which the combination of the two has a global pressure minimum.
The rarefaction-induced pressure field in Fig.~\ref{fig:CompressionVsRarefaction}(c) does not show a particularly localized focal region, but rather a broad amplification of the wave traveling through the droplet core.
A maximum amplification of 1.1 times the incoming wave PNP is achieved at the location corresponding to the global pressure minimum. 
On the other hand, surprisingly, the compression-induced solution in Fig.~\ref{fig:CompressionVsRarefaction}(d) shows a highly localized negative pressure peak with a maximum amplification factor of 1.05 with respect to the incoming wave PNP. The location of this localized negative pressure peak on the distal side of the droplet matches well with the minimum pressure achieved in the combined solution, suggesting that the compression phase dictates the location of the latter in addition to contributing to its absolute value.

Since the acoustic impedance $Z = \rho c$ of PFP is lower than water, tension generation cannot be attributed to the reflection of the monopolar compression wave at the internal boundary of the droplet. Instead, it can be seen from animations (see Supplementary Material~\cite{noauthor_notitle_nodate}) that the monopolar compression wave, once transmitted into the droplet core, converges and, at its focus, develops a negative tail that is then reflected at the rear boundary and refocused at the second focal position in the proximal side of the droplet.
Here, the negative pressure generated by the compression peak combines with the amplified negative pressure obtained from the transmission of the rarefaction phase, thus generating a global minimum throughout the whole excitation period.
We believe that a so-called \emph{Gouy phase shift} happens while the compression wave crosses both focal points.
Louis Georges Gouy first reported this effect in 1890 \cite{gouy_sur_1890} for electromagnetic waves, but this principle can be applied to any wave subjected to transverse spatial confinement \cite{kaltenecker_gouy_2016} and has recently been experimentally verified for focusing acoustic waves \cite{lee_origin_2020}.
For a three-dimensional focusing beam, the Gouy phase shift consists of an additional variation in phase of $ -\pi $ with respect to a plane wave traveling in the same medium, gradually acquired as the wave crosses the focal point \cite{tyc_gouy_2012} and corresponding to a sign inversion. 

\begin{figure}[th]
    \includegraphics[width = 0.5\columnwidth]{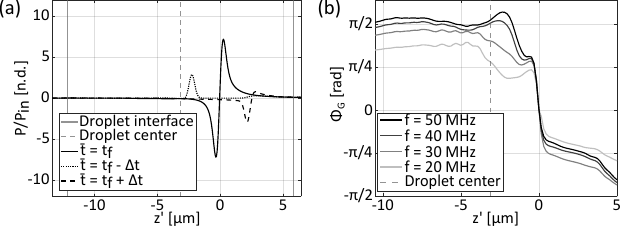}
    \caption{\label{fig:PhaseShift} (a)~Selected instants of the pressure field computed at the centerline of a droplet due to interaction with a plane wave having a purely compressive Gaussian temporal profile. The time at which the weave reaches the focal point is $t_f$, while $\Delta t = 5.6\ \si{\nano\second}$. (b)~Calculated phase shift of different frequency components of a Gaussian wave focusing inside a 9-\si{\micro\meter}-radius droplet.
    In both cases, $z^{\prime} = z - z_f$ represents the distance from the droplet focal point $z_f$. }
\end{figure}

Due to the high longitudinal spatial confinement within the droplet, classical modeling of the Gouy phase shift, which relies on the fact that the wave travels in an infinite medium \cite{feng_physical_2001, zhu_direct_2007, tyc_gouy_2012, kaltenecker_gouy_2016}, cannot be directly applied.
To assess the phase shift of monochromatic plane waves with different frequencies, we performed a numerical simulation in which a plane wave with a Gaussian temporal profile $ f(t) = e^{(t - \mu)/2\sigma} $ ($ \mu = 3.3 \; \si{\nano\second}$, $\sigma = 0.55 \; \si{\nano\second} $) is transmitted and focused in a 9-\si{\micro\meter}-radius droplet. Three snapshots are represented in Fig.~\ref{fig:PhaseShift}(a) at the times $t_f \pm 5.6 \;\si{\nano\second}$, with $t_f$ being the time at which the wave reaches the focal point.
Most of the harmonics of interest contained in the focusing wave can be analyzed by performing a Fourier transform of the signal inside the droplet core at different time instants (i.e., different positions of the propagating wave in the droplet core). For each selected instant and each of the constitutive Fourier harmonics, the phase acquired by a plane wave traveling in the same medium, $ \Phi_{n}^{ref} = 2\pi f_n \Delta t = 2\pi \lambda_n \Delta z $, can be deducted to estimate the Gouy phase, following a procedure similar to the one found in Zhu \emph{et al.}~\cite{zhu_direct_2007}. Here, $ f_n $ and $ \lambda_n $ represent the frequency and wavelength of the $n$-th harmonic, and $ \Delta t = \Delta z / c_2 $ represents the time needed for the wave to travel a distance $ \Delta z $ in the droplet. 
The Gouy phase is set to equal zero at the instant the wave reaches its focus, $ \Phi_G(z^{\prime} = 0) = 0 $.
We calculated a phase shift of around $ \pi $ across the focal point, shown in  Fig.~\ref{fig:PhaseShift}(b), for the harmonic components of the incoming wave with frequency $ f \simeq 20~\si{\mega\hertz}$ and above. Calculating this for lower frequencies is not possible due to the Fourier transform resolution limitation $ \Delta f = c_2/2R \simeq 20~\si{\mega\hertz} $. 
Moreover, monochromatic plane waves with frequencies lower than 20~\si{\mega\hertz} (i.e. wavelengths larger than the droplet diameter), do not show focusing as predicted by geometrical acoustics but present a maximum at the droplet center, as the focusing wave interacts with its reflection at the droplet wall within the span of its wavelength.

\begin{figure}[tb]
        \includegraphics[width = 0.5\columnwidth]{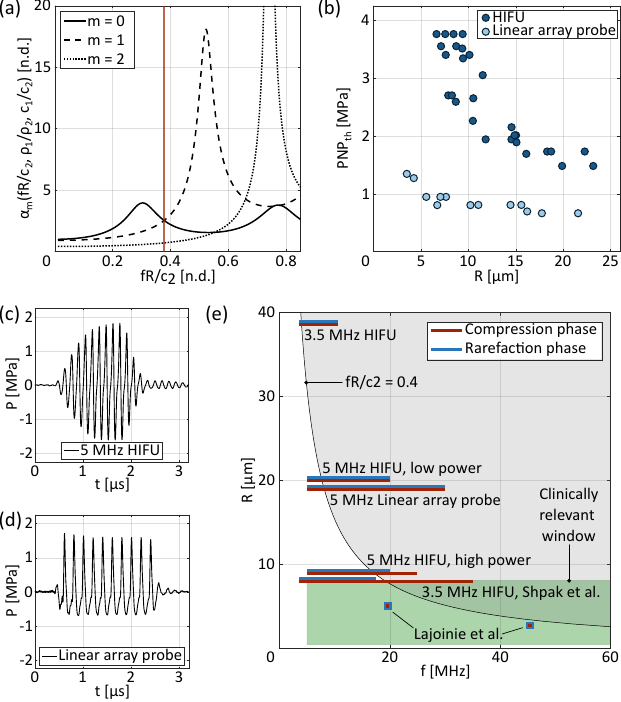}
        \caption{\label{fig:modeAmplificationAndPhaseShiftoccurrence} (a)~Amplification factor for the acoustic modes of order $m$ of a PFP droplet surrounded by water. The red line represents the point at which the amplification of the mode $m = 1 $ is higher than the amplification for the zeroth order mode. 
        (b)~Vaporization PNP threshold of PFP droplets experimentally observed for two different ultrasound transducers as a function of droplet radius. The pressure measurements have an accuracy of $\pm 10\%$, while the error on the radius measurements is estimated to be $\pm 0.32$ \si{\micro\meter}. 
        (c)-(d) Waveforms of the 5-\si{\mega\hertz} HIFU and the linear array probe driven at 5-\si{\mega\hertz}, respectively.        
        (e)~Harmonic content of the rarefaction and compression phase of the waves employed in this study and some selected works from literature plotted on the $(f, R)$ space. The gray area represents the $(f, R)$ value-pairs in the ``phase-shift" regime. The green area represents the clinically relevant operating window, considering the maximum size allowed for intravenous agents and the minimum frequency typically employed for ultrasound diagnostics \cite{borden_reverse_2018}.}
\end{figure}

To better describe this behavior, acoustic mode decomposition can be employed.
From Eq.~(\ref{eq:transmittedPressure}) it is clear that every harmonic component of the driving wave excites an infinite number of acoustic modes of the droplet, represented by Bessel functions of the first kind multiplied by Legendre polynomials.
Every acoustic mode is multiplied by a gain factor $ \tilde{\alpha}_m $ which, in dimensionless variables, depends on the ratio between density $ \rho_{1}/\rho_2 $ and speed of sound $c_1/c_2$ of the droplet and surrounding medium, the order of the considered mode $ m $, and the dimensionless droplet radius $\widetilde{R} = fR/c_2$. The dependency on the harmonic index $ n $ is incorporated in the generic frequency, $ f $. 
Assuming fixed material properties, the problem can be compactly described by $ \widetilde{R}$, which includes the combined effect of the frequency of the driving wave and the droplet radius.
Curves of the amplification factors $\tilde{\alpha}_m$ for the acoustic modes of order $ m $ are presented in Fig.~\ref{fig:modeAmplificationAndPhaseShiftoccurrence}(a), and are calculated using the physical properties for water and PFP.
    
Every mode presents multiple resonance frequencies in which the amplification has a local maximum, as depicted in Fig.~\ref{fig:modeAmplificationAndPhaseShiftoccurrence}(a) for the mode $m = 0$.
For a monochromatic wave, the full set of amplification factors required to calculate the theoretical solution can be obtained by intersecting a vertical line $ \widetilde{R} = \bar{f}\overline{R}/c_2$ for fixed values of $R$ and $f$, with the curves of the different $ \tilde{\alpha}_m $.
Modes with $ m \geq 1 $ share the property of being equal to zero at the droplet center, where instead the mode $ m = 0 $ presents a global maximum due to the shape of the spherical Bessel functions.
Therefore, if the zeroth mode is the most excited, focusing of the wave in the classical sense, i.e.,\ at the focal point predicted by geometrical acoustics, is not clearly visible, and the droplet will be in an ``acoustic resonance'' regime, similar to what is described by Lajoinie \emph{et al.}~\cite{lajoinie_high-frequency_2021}.
However, if higher modes are excited, two main focal points will be visible, one on the distal and one on the proximal side of the droplet due to successive reflection at the droplet wall. The droplet will be in a ``phase-shift'' regime, with wave sign inversion occurring as it crosses the focal points (distal and proximal). This can be interpreted as the ``superharmonic focusing'' effect reported by Shpak \emph{et al.}~\cite{shpak_acoustic_2014}.
The transition between the two regimes is gradual, but a conventional limit can be fixed at the $\widetilde{R}$ value where the gain for the mode $ m = 1 $ overcomes the one for $ m = 0 $, corresponding to $ \widetilde{R} \simeq 0.4 $, which is consistent with $f \geq c_2/2R $ estimated for the effect of droplet confinement on wave focusing. The limit is represented in Fig.~\ref{fig:modeAmplificationAndPhaseShiftoccurrence}(a) as a vertical red line.

The theory is consistent with the experimental results presented beforehand. The harmonic content of the 5-\si{\mega\hertz} wave presented in Fig. \ref{fig:CompressionVsRarefaction}(b) shows a negligible rarefaction component above 20~\si{\mega\hertz}, which corresponds to a dimensionless radius of $ \overline{R}_{\mathrm{rar}} \backsim 0.44 $ for a 9-\si{\micro\meter}-radius droplet. As a result, the energy content of the rarefaction phase will mostly excite the mode $m = 0$. The compression phase instead presents energy in the fifth harmonic (25~\si{\mega\hertz}), corresponding to a value $ \overline{R}_{\mathrm{comp}} \backsim 0.55 $ very close to the $m$ = 1 resonance peak, thus explaining the localization of the nucleation spots.
Similarly, the 3.5-\si{\mega\hertz} wave employed for the sonication of 38-\si{\micro\meter}-radius droplets contains significant energy in the first harmonic (both for the compression and rarefaction phase), corresponding to $\overline{R}_{\mathrm{comp}} \backsim 0.33$. Despite the calculated position of the pressure minimum being slightly away from the droplet center, most likely due to the effect of higher harmonics in the driving wave, the nucleation event is dominated by the resonance regime, as demonstrated by the nucleation map in Fig.~\ref{fig:setupPressureMap}(c). 

A direct implication of the wave's positive pressure contribution to enhance tension generation is a significant reduction of the vaporization threshold $\text{PNP}_\mathrm{th}$, as shown in Fig.~\ref{fig:modeAmplificationAndPhaseShiftoccurrence}(b). This parameter is defined by extracting the PNP of the driving wave for which single droplet vaporization is first observed optically, obtained by gradually increasing the energy supplied to the transducer.
The apparent distinction between the 5-\si{\mega\hertz} HIFU and the linear array probe in the vaporization threshold can be attributed to the different harmonic content of the two excitation waves, with the HIFU wave in Fig.~\ref{fig:modeAmplificationAndPhaseShiftoccurrence}(c) being much less distorted than the one driven by the linear array probe in Fig.~\ref{fig:modeAmplificationAndPhaseShiftoccurrence}(d) (see also Supplementary Material \cite{noauthor_notitle_nodate}). Wave distortion enables significant tension generation with relatively low PNP, facilitated by the phase-shift of the compression part of the driving wave.

Once the critical value of the dimensionless radius has been set, the function, $ \widetilde{R} = 0.4 $, can be plotted on the frequency-radius space, thus delimiting the ``resonance'' and the ``phase-shift'' regime on the $(f, R)$ space. The resulting graph is presented in Fig.~\ref{fig:modeAmplificationAndPhaseShiftoccurrence}(e), where the gray-shaded area represents the necessary conditions for phase shift occurrence within the droplet core.
The harmonic content of the compression and rarefaction phases is reported in correspondence with the radius and frequency tested in our experiments and in selected past studies \cite{shpak_acoustic_2014, lajoinie_high-frequency_2021}. This graph provides an intuitive way to assess the possibility of ``resonance" and ``phase-shift" by looking at the location of the harmonic content in the $(f, R)$ space. 
In general, the presence of harmonics in the ``phase-shift" regime indicates the opportunity to achieve elevated tension, either due to the focusing of the rarefaction part of the wave or the phase shift of the compression one. A more accurate prediction would however require an assessment of the occurrence of resonances by inspecting the $ \widetilde{R}$ values for every harmonic of the driving wave.

\emph{Conclusions.} The theory described here helps demonstrate the significance of the compression phase of a distorted ultrasound wave in promoting ADV. The main outcome lies in the finding that for a given radius, harmonics with a frequency $ f > 0.4c_2/R $ present a pronounced phase shift while crossing the focal point of the PFP droplet resulting in sign reversal of the waveform.
The distortion of a wave due to nonlinear propagation in the medium causes the high-frequency harmonic content to concentrate in the compression phase of the ultrasound, promoting the occurrence of Gouy phase shift at the focus and the generation of a localized tension region. The proposed scaling using the dimensionless radius allows extending the range of droplet sizes while maintaining a similar pressure field with the practically relevant solutions (i.e., green area in Fig. \ref{fig:modeAmplificationAndPhaseShiftoccurrence}(e)), potentially solving the inherent technological limitation of optical observation of sub-micron droplet dynamics. 

Our results suggest that describing the acoustic vaporization threshold of a droplet solely by the applied PNP of the incoming wave is insufficient.
The full extent of the wave spectrum is rarely taken into account despite its importance in determining the maximum tension generated in the droplet core, and this could be a reason behind the discrepancies between the vaporization thresholds reported by prior ADV studies \cite{aliabouzar_acoustic_2018, wu_investigation_2021}.

The occurrence of the phase shift offers the opportunity to optimize the acoustic driving for micro- and nanodroplets by designing the shape of the driving wave. Distorted waves with a weak rarefaction phase and a strong peak positive pressure can generate the required tension in the droplet core without affecting the surrounding medium, improving the safety of ADV while still achieving the desired therapeutic effects.

\emph{Acknowledgments.} This work was funded by the Swiss National Science Foundation (SNSF project number 200021\_200567). We thank Prof.\ Inge Herrmann for lending us the $ 3.5$-$\si{\mega\hertz} $ HIFU transducer.

\bibliographystyle{apsrev4-2}
\bibliography{main.bib}

\end{document}


\title{Supplementary Material \\ Positive pressure matters in acoustic droplet vaporization}

\author{Samuele Fiorini}
\author{Anunay Prasanna}
\author{Gazendra Shakya}
\author{Marco Cattaneo}
\author{Outi Supponen}

\affiliation{Institute of Fluid Dynamics, ETH Zürich, Zürich, Switzerland}

\maketitle

\section{Theory}
The theoretical model used in the present Letter closely follows the modeling proposed by Shpak \emph{et al.}~\cite{shpak_acoustic_2014} and Anderson \cite{anderson_sound_1950, feuillade_anderson_1999}. The complete derivation is provided here for the benefit of the reader.

Due to its small size with respect to the wavelength of the driving ultrasound, nonlinear propagation within the droplet can be neglected, and the resulting pressure field inside the liquid core can be modeled using the linear wave equation~\cite{shpak_acoustic_2014},
\begin{equation}
    c^2\nabla^2p -\partial^2 p/\partial t^2 = 0.
    \label{eq:wave}
\end{equation}
The experimentally recorded pressure waveform is used as an input for the theoretical calculation, approximating the excitation signal as a plane wave because of the small size of the droplet with respect to the transducer's focal region. The signal can be expanded in the Fourier series, obtaining an expression of the form \cite{shpak_acoustic_2014}:
    \begin{equation}
    p_f(t) = \Re \left(\sum_{n = 0}^{\infty}a_ne^{i(n\omega t + \phi_n)}\right),
    \end{equation}
    where $ a_n $ is the amplitude, $ \phi_n $ the phase of the $n$-th harmonic component of the ultrasound wave and $ \omega $ is the fundamental angular frequency.  
Due to the linearity of the wave equation, the solution for an arbitrary periodic waveform can be expressed as the sum of the solution for each harmonic component, $n$.
The theoretical solution for a plane wave impinging on a droplet is therefore only presented for a monochromatic plane wave, $ \widetilde{p}_{i}(z, t) = a e^{i(\omega t - kz + \phi)}$, where $a,\; \omega,\; k \; \text{and} \; \phi$ are the wave amplitude, angular frequency, wavenumber and initial phase of the wave, respectively. 

\begin{figure}[h]
    \centering
    \includegraphics[width = 0.4\textwidth]{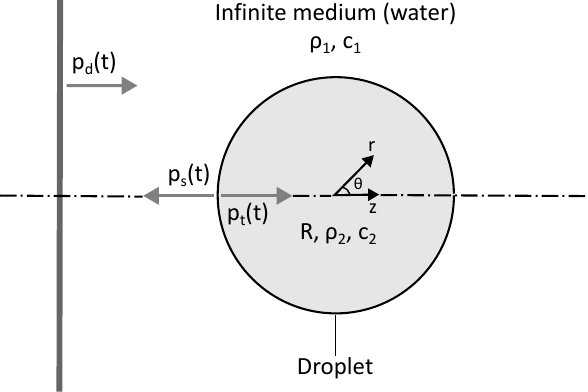}
    \caption{Illustration of an acoustic wave interacting with a spherical droplet in an infinite medium.}
    \label{fig:theoryschematic}
\end{figure}

Consider a liquid sphere of radius $ R $ with density $ \rho_2 $ and sound speed $ c_2 $, immersed in a homogeneous fluid with density and sound speed $ \rho_1 $, $ c_1 $.
The origin of the spherical coordinate system ($ r, \theta, \varphi $) is set at the center of the droplet, and the plane wave is traveling in the $z$-direction. The solution is assumed to be of the form 
\begin{equation}
    \widetilde{p}_{\eta}(r, \theta, \varphi, t) = a\widetilde{\Psi}_{\eta}(r, \theta, \varphi)e^{i \omega t + \phi},
    \label{eq:generalSolutionVariableSeparation}
\end{equation}
where $ \widetilde{\Psi}_{\eta}(r, \theta, \varphi) $ is the solution of the Helmholtz equation $ \nabla^2\widetilde{\Psi}_{\eta} + k^2\widetilde{\Psi}_{\eta} = 0 $ \cite{kinsler_fundamentals_2000}, the wavenumber $ k = \omega/c $ is obtained by substituting Eq.~(\ref{eq:generalSolutionVariableSeparation}) into Eq.~(\ref{eq:wave}), and $\eta = d, s, t $ denote the driving, scattered and transmitted wave respectively.. 
A schematic is presented in Fig. \ref{fig:theoryschematic}.
Due to the symmetry of the problem, only the dependence on the polar angle $ \theta $ and the radial direction $ r $ are considered.
The distorted driving wave can be expanded in spherical harmonics~\cite{strutt_theory_2011},
 \begin{equation}
    \widetilde{\Psi}_d = e^{-ikr\cos\theta} = \sum_{m = 0}^{\infty} (-i)^m(2m + 1)j_m(k_1r)P_m(\cos{\theta})
    \label{eq:incomingWaveSphericalHarmonics}
\end{equation}
where $ m $ represents the index of the acoustic mode of the system under examination, $ k_1 = \omega/c_1 $ the wavenumber in the undisturbed medium, $ j_m(k_1r) $ the spherical Bessel function of the first kind of order $ m $ \cite{abramowitz_handbook_1972}, and $ P_m $ the Legendre polynomial of order $m$ \cite{kinsler_fundamentals_2000}.
For the region outside the droplet, the complete solution consists of the sum of the driving wave $ \widetilde{\Psi}_d  $, and a scattered wave $ \widetilde{\Psi}_s $,
\begin{equation}
    \widetilde{\Psi}_s = \sum_{m = 0}^{\infty} \widetilde{\beta}_m (j_m(k_1r) - i y_m(k_1r))P_m(\cos{\theta}),
    \label{eq:scatteredWave}
\end{equation}
where $ y_m(k_1r) $ is the spherical Bessel function of the second kind \cite{abramowitz_handbook_1972} and $ \widetilde{\beta}_m $ is an unknown coefficient.
The subtraction between spherical Bessel functions, $(j_m - iy_m)$ in Eq.~(\ref{eq:scatteredWave}), ensures that the Sommerfield radiation condition is respected. From a physical point of view, this choice can be justified by looking at the asymptotic expansion of the Bessel functions for $ kr \xrightarrow[]{} \infty $,
   \begin{equation}
        j_m(kr) - i y_m(kr) \xrightarrow[]{kr \to \infty} (-i)^m\frac{1}{kr}e^{-ikr}
    \end{equation}
which, multiplied by the time-dependent part of the complete solution $ e^{i\omega t} $, gives the expression of a spherical wave traveling away from the system's origin, as one would expect from a wave scattered by an object. 
Inside the droplet core, only the transmitted wave is present and can be written as,
\begin{equation}
     \widetilde{\Psi}_t = \sum_{m = 0}^{\infty} \widetilde{\alpha}_m j_m(k_2r)P_m(\cos{\theta}),
     \label{eq:transmittedWave}
\end{equation}
where $k_2 = \omega/c_2 $ is the wavenumber inside the droplet and $ \widetilde{\alpha}_m $ is an unknown coefficient, to be determined by satisfying the boundary conditions. 
In this case, the pressure field is represented as a sum of spherical Bessel function of the first kind only.
This can be interpreted as a sum of inward and outward traveling waves described before, which implies that $ j_m(k_2r) $ physically represents a standing wave inside the droplet core.

At the sphere's boundary, $ r = R\ \forall (\theta, t) $, both the acoustic pressure and the radial particle velocity must be continuous. Therefore, the two boundary conditions can be written as
\begin{equation}
    \label{eq:boundaryConditionsPressure}
     \widetilde{p}_d(R) + \widetilde{p}_s(R) = \widetilde{p}_t(R)
\end{equation}
\begin{equation}
    \label{eq:boundaryConditionsVelocity}
    \frac{1}{\rho_1 c_1}\eval{\left( \frac{\partial \widetilde{p_i}(r)}{\partial (k_1 r)} + \frac{\partial \widetilde{p_s}(r)}{\partial (k_1 r)}\right)}_{r = R} = \frac{1}{\rho_2 c_2}\eval{\frac{\partial \widetilde{p_t}(r)}{\partial (k_2 r)}}_{r = R},
\end{equation}
where the dependency on time and the polar angle in Eq.~(\ref{eq:boundaryConditionsPressure}) and (\ref{eq:boundaryConditionsVelocity}) have not been reported for ease of notation.
Inserting Eqs.~(\ref{eq:incomingWaveSphericalHarmonics}), (\ref{eq:scatteredWave}), and (\ref{eq:transmittedWave}) into Eqs.~(\ref{eq:boundaryConditionsPressure}) and (\ref{eq:boundaryConditionsVelocity}), the unknown coefficients $ \widetilde{\alpha} $ and $ \widetilde{\beta} $ can be determined,
\begin{equation}
    \widetilde{\alpha}_m = (-i)^m(2m + 1)\frac{j_m(k_1R)h_m^{\prime(2)}(k_1R) - h_m^{(2)}(k_1R)j_m^{\prime}(k_1R)}{j_m(k_2R)h_m^{\prime(2)}(k_1R) - \frac{\rho_1c_1}{\rho_2c_2}h_m^{(2)}(k_1R)j_m^{\prime}(k_2R)}
    \label{eq:amplficationFactors}
\end{equation}
\begin{equation}
    \widetilde{\beta}_m = (-i)^m(2m + 1)\frac{\rho_1c_1j_m(k_1R)j_m^{\prime}(k_2R) - \rho_2c_2j_m(k_2R)j_m^{\prime}(k_1R)}{\rho_2c_2j_m(k_2R)h_m^{\prime(2)}(k_1R) - \rho_1c_1h_m^{(2)}(k_1R)j_m^{\prime}(k_2R)},
    \label{eq:amplficationFactorsScattered}
\end{equation}
where $ h_m(kr) = j_m(kr) - iy_m(kr) $ is the spherical Hankel function of second kind of order $ m $. Interestingly, if the ratio between the densities and sound speed of the materials is fixed, the coefficients $ \widetilde{\alpha}_m $ and $ \widetilde{\beta}_m $ are completely determined by the order $ m $, and the dimensionless radius $ \widetilde{R} = fR/c_2 $. 

The expression for the full solution of the pressure inside the droplet core is finally obtained by summing the solution for every harmonic component, $n$, of the driving wave and extracting the real part:
\begin{equation}
    p_{t}(r, \theta, t) = \Re \left[ \sum_{n = 0}^{\infty} \sum_{m = 0}^{\infty} a_ne^{in\omega t + \phi_n}\widetilde{\alpha}_{m,n}j_m(nk_2r)P_m(\cos{\theta})\right]
    \label{eq:transmittedPressure}
\end{equation}
In similar fashion, the expression for the driving and scattered pressure wave can be summed together to obtain the pressure field outside the droplet:
\begin{equation}
    p_{\rm ext}(r, \theta, t) = \Re \left\{ \sum_{n = 0}^{\infty} \sum_{m = 0}^{\infty} a_ne^{in\omega t + \phi_n}[(-i)^m(2m + 1)j_m(nk_1r) + \widetilde{\beta}_{m,n}h_m^{(2)}(nk_1r)]P_m(\cos{\theta})\right\}.
    \label{eq:outsidePressure}
\end{equation}
 
\section{Numerical simulation}~
The transmission of the ultrasound wave inside the droplet has been simulated using the k-Wave MATLAB Toolbox \cite{treeby_k-wave_2010}. Due to the geometry of the problem, axial axisymmetry has been considered in the direction of wave propagation. The droplet was modeled as a discontinuity in density and sound speed positioned at the center of the domain. Properties of perfluoropentane \cite{hallewell_properties_2010} and water \cite{wagner_iapws_2002} has been used to model the droplet and the surrounding fluid. The data has been interpolated in order to get the properties at 37\si{\celsius} and 40\si{\celsius}, depending on the simulated experimental conditions.  The parameters are summarized in Table \ref{tab:simulationParameter}. 

\begin{table}[h]
\begin{tabular}{l|cccc}
 & Water (37\si{\celsius}) & PFP (37\si{\celsius}) & Water (40\si{\celsius}) & PFP (40\si{\celsius}) \\ \hline\hline
Density [\si{\kilo\gram\per\cubic\meter}] & 993 & 1571 & 992 & 1560 \\
Speed of sound [\si{\meter\per\second}] & 1524 & 406 & 1529 & 397
\end{tabular}
\label{tab:simulationParameter}
\caption{Physical properties of water and perfluoropentane (PFP) used in the k-Wave simulations. The values are obtained from interpolation of the saturation curves provided in \cite{hallewell_properties_2010, wagner_iapws_2002}}.
\end{table}

A velocity source in the direction of propagation of the wave has been used to create a plane wave travelling in the computational domain. To reproduce the experimental conditions, the pressure signal recorded at the focal point of the transducer was used as an input time series for the velocity source, after an appropriate scaling following the acoustic relation for plane waves \cite{kinsler_fundamentals_2000}: 
\begin{equation}
\tilde{u}_{\pm}(t) = \pm\frac{\tilde{p}_{\pm}(t)}{\rho_1 c_1},
\end{equation}

where $\tilde{u}_{\pm}$ is the particle velocity of the wave, $\tilde{p}_{\pm}$ the acoustic pressure, $ \rho_1 $ and $ c_ 1 $ the density and speed of sound of the medium surrounding the droplet. The superscripts "$+$" and "$-$" indicate a wave traveling in the positive and negative direction of the propagation axis. 

\begin{figure}[h]
    \centering
    \includegraphics[width = 0.35\textwidth]{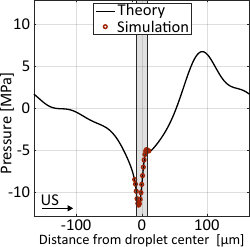}
    \caption{Comparison between the simulated and theoretically calculated pressure field on the symmetry axis for a 9-\si{\micro\meter} radius droplet generated by the interaction with the experimentally measured 5-\si{\mega\hertz} HIFU pressure signal. The shaded area represent the droplet position.}
    \label{fig:theorySimulationComparison}
\end{figure}

The simulation results have been used to validate the theoretical model described in the previous section with very good agreement, as showed in Fig. \ref{fig:theorySimulationComparison}.

\newpage

\section{Harmonic content of a 5-MHz HIFU (low power) and a 5-MHz linear array probe pressure wave}

\begin{figure}[h]
    \centering
    \includegraphics[width = 0.8\textwidth]{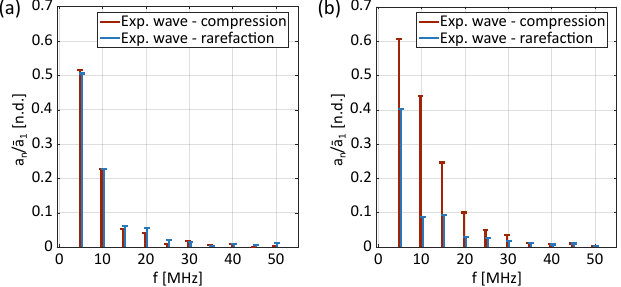}
    \caption{(a) Harmonic content of the compression and rarefaction phase of the experimentally measured pressure wave from a 5-MHz HIFU transducer (driven at low power) and (b) a 5-MHz linear array probe. The values on the y-axis are normalized by the coefficient of the fundamental harmonic of the complete wave.}
    \label{fig:extendedHarmonicContent}
\end{figure}

\bibliographystyle{apsrev4-2}
\bibliography{supp.bib}